\documentstyle[emulateapj]{article}

\def\gsnr{G328.4+0.2}
\def\newsrc{AX J1555.5$-$5318}
\def\mshsnr{MSH 15$-$5{\sl 7}}

\def\eo{{\it Einstein Observatory}}
\def\rosat{{\it ROSAT}}
\def\asca{{\it ASCA}}

\def\myarcmin{^\prime\mskip-5mu.}

\def\lsim{\hbox{\raise.35ex\rlap{$<$}\lower.6ex\hbox{$\sim$}\ }}
\def\gsim{\hbox{\raise.35ex\rlap{$>$}\lower.6ex\hbox{$\sim$}\ }}

\submitted{Received 2000 March 15; accepted 2000 May 8}
\accepted{To Appear in the Astrophysical Journal, Volume 542, 10 October 2000}

\lefthead{X-rays from \gsnr}
\righthead{Hughes, Slane, \& Plucinsky}

\begin{document}

\title{Discovery of X-ray Emission from \gsnr, a Crab-Like Supernova 
Remnant}

\author{
John P.~Hughes\altaffilmark{1}}
\affil{Department of Physics and Astronomy, Rutgers University, 136 
 Frelinghuysen Road, Piscataway NJ 08854-8019; jph@physics.rutgers.edu
}
\altaffiltext{1}{
Also Service d'Astrophysique, L'Orme des Merisiers, 
CEA-Saclay, 91191 Gif-sur-Yvette Cedex France
}

\author{Patrick O.~Slane and Paul P.~Plucinsky}
\affil{Harvard-Smithsonian Center for Astrophysics, 60 Garden Street,
Cambridge MA 02138  
}

\begin{abstract}

\gsnr\ is a moderately small ($5^\prime\times5^\prime$) Galactic radio
supernova remnant (SNR) at a distance of at least 17 kpc that has
been long suggested to be Crab-like. Here we report on the detection
with \asca\ of the X-ray emission from the SNR.  The X-ray source is
faint with an observed flux of $(6.0\pm0.8)\times 10^{-13}$ erg
s$^{-1}$ cm$^{-2}$ over the 2--10 keV band. The emission is heavily
cut-off at low energies and no flux is detected below 2 keV.  Spectral
analysis confirms that the column density to the source is indeed
large, $N_{\rm H} \sim 10^{23}$ atoms cm$^{-2}$, and consistent with
the total column density of hydrogen through the Galaxy at this
position.  Good fits to the spectrum can be obtained for either
thermal plasma or nonthermal power-law models, although the lack of
detected line emission as well as other evidence argues against the
former interpretation.  The power-law index we find, $\alpha_P =
2.9^{+0.9}_{-0.8}$, is consistent with other Crab-like SNRs. In the
radio band \gsnr\ is nearly as luminous as the Crab Nebula, yet in the
X-ray band luminosity it is some 70 times fainter. Nevertheless its
inferred soft X-ray band luminosity is greater than all but the
brightest pulsar-powered synchrotron nebulae and implies that \gsnr\
contains a rapidly spinning, as yet undetected, pulsar that is losing
energy at a rate of $\sim$$10^{38}$ erg s$^{-1}$.

\end{abstract}


\section{INTRODUCTION}

Only about one-quarter of all cataloged Galactic radio-emitting
supernova remnants (SNRs) (Green 1998) are known to emit X-rays based
largely on observations by the \eo\ and \rosat.  Some, and perhaps
even many, of the non-detections are likely a result of the absorption
of the typically soft (i.e., $<$2 keV) X-ray emission from SNRs by the
large column density of gas and dust in the Galactic plane.  However,
young remnants, like Cassiopeia A, Tycho, and Kepler that show
significant thermal emission above 2 keV, and pulsar-powered
synchrotron nebulae, like the Crab Nebula with hard, featureless
power-law spectra, should be able to shine through the interstellar
medium.  In fact the entire Galaxy is effectively transparent to
X-rays with energies above about 4 keV.  As part of an effort to
identify and study such SNRs, we have targeted a sample of small
diameter Galactic remnants with the {\it Advanced Satellite for
Cosmology and Astrophysics} (\asca), which has high sensitivity and
moderate imaging capabilities in the hard X-ray band.  Here we report
on the discovery of X-ray emission from one of the remnants in the
sample, \gsnr.

\gsnr\ (\mshsnr) has been cataloged as a radio SNR for quite some time
(see, e.g., Clark \& Caswell 1976) and, because of its flat radio
spectrum, $S_{\nu} \propto \nu^{-0.24}$ (Milne 1979), and
``filled-center'' or plerionic morphology (Caswell et al.~1980;
Whiteoak \& Green 1996), it has long been considered as a possible
Crab-like remnant.  Its distance, based on \ion{H}{1} absorption
studies (Caswell et al.~1975; Gaensler, Dickel, \& Green 2000
hereafter GDG), is rather large, 17 kpc or more, corrected to
currently accepted values for the rotation curve and distance to the
Galactic center.  No pulsar has yet been detected (Kaspi et al.~1996)
and previous searches for X-ray emission have been unsuccessful
(Wilson 1986).

\section{ANALYSIS}

We observed \gsnr\ with \asca\ (Tanaka, Inoue, \& Holt 1994) on 17
March 1998.  The solid-state imaging spectrometer (SIS), which
comprises two separate arrays of CCD cameras, was operated in 2-CCD
mode. The gas imaging spectrometer (GIS), consisting of two gas
scintillation proportional counters, was operated in PH mode with the
re-assignment of 2 bits of telemetry from pulse-height (compressing
the spectra from 1024 channels to 256) to timing.  This, in fact,
resulted in no loss of spectral resolution since the GIS spectra are
re-binned to this resolution before spectral analysis
anyway. Rise-time information was used for suppressing background
events.  The data were screened using standard criteria that resulted
in a combined exposure of 32900 s for the SIS detectors and 38950 s
for the GIS detectors.  The data from each detector were separately
reduced and like observations were combined.

\begin{figure*}[t]
\plotfiddle{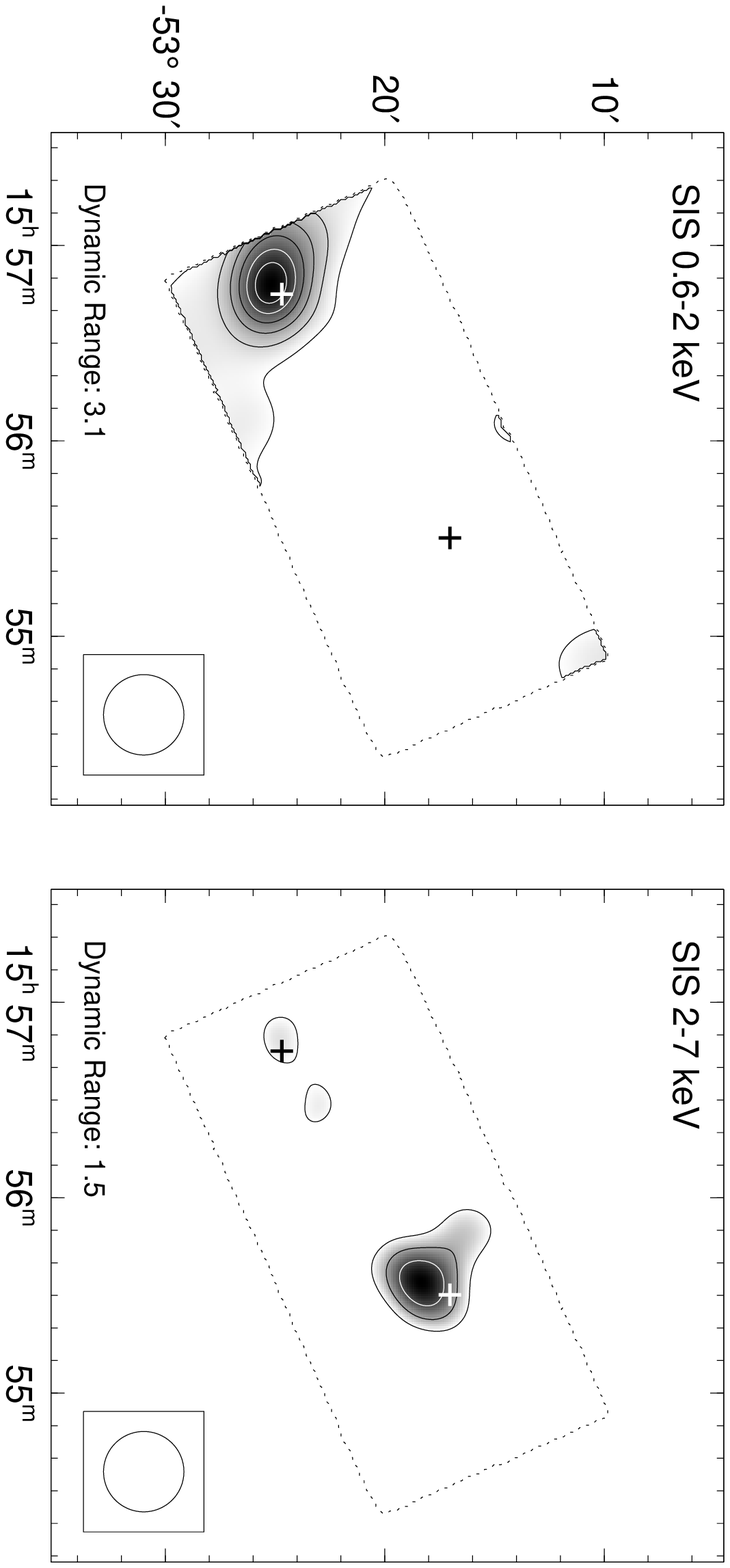}{4in}{90}{75}{75}{+300}{-30}
\vskip -0.75truein
\figcaption[f1.ps]{(a) X-ray image of the SNR \gsnr\ and environs from
a 32900 s observation with the \asca\ SIS in soft and hard X-ray bands
as indicated.  Plus signs denote the location of the radio SNR (the
source to the west) and a previously known soft X-ray source.  The
data were smoothed with a 1$^\prime$ ($\sigma$) gaussian. In the soft
band image, the greyscale runs linearly from $1.87 \times 10^{-4}
\rm\, SIS\, counts \, s^{-1}\, arcmin^{-2}$ (3 $\sigma$ above
background) to $5.76 \times 10^{-4} \rm\, SIS\, counts \, s^{-1}\,
arcmin^{-2}$ (the peak emission). In the hard band image, the
greyscale runs linearly from $3.00 \times 10^{-4} \rm\, SIS\, counts
\, s^{-1}\, arcmin^{-2}$ (3 $\sigma$ above background) to $4.55 \times
10^{-4} \rm\, SIS\, counts \, s^{-1}\, arcmin^{-2}$ (the peak
emission).  In each panel contours are plotted at linear intervals
from the minimum value to 90\% of the peak.  The dotted contour
indicates the edge of the detector.  The approximate size of the
point-spread-function is indicated in the lower right.}
\plotfiddle{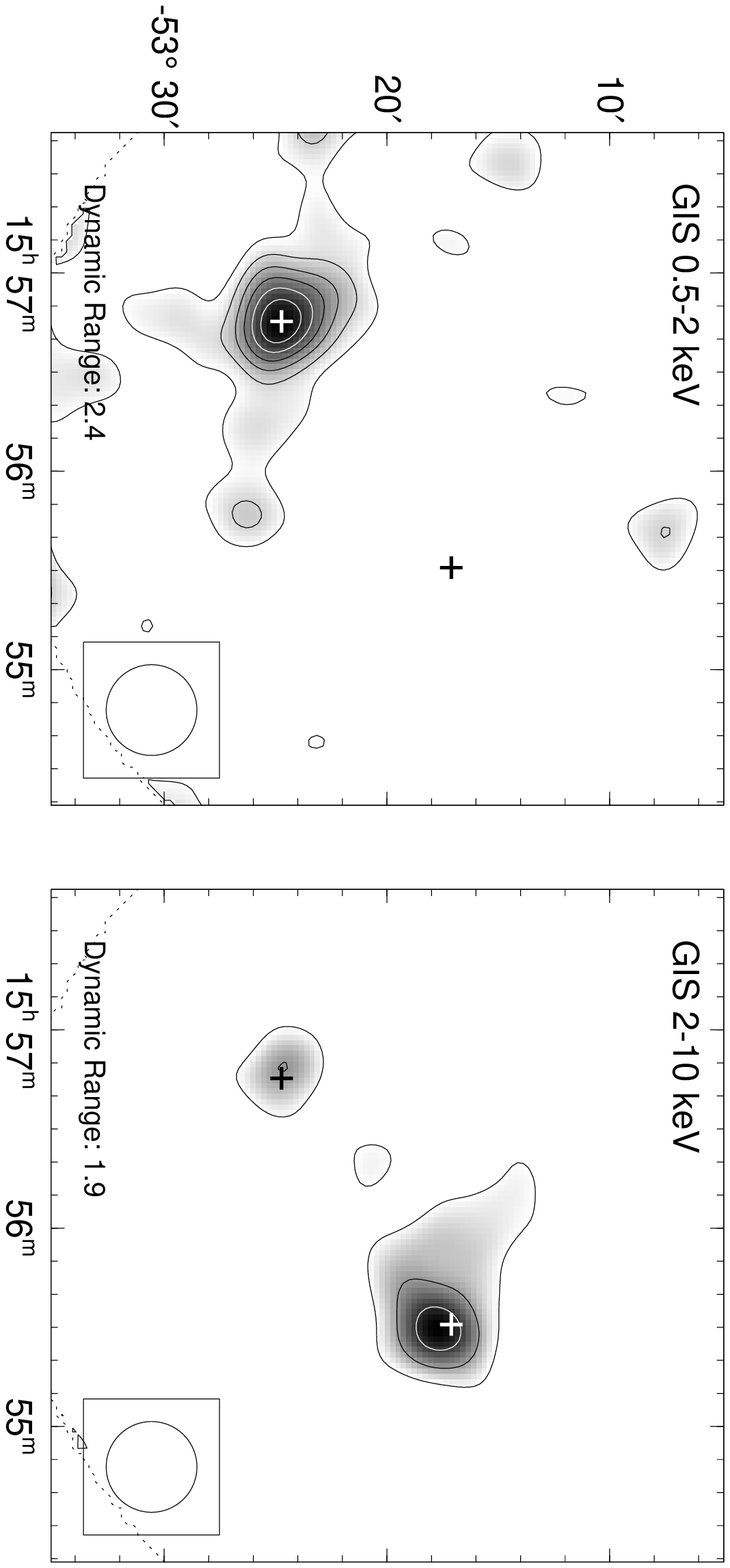}{4in}{90}{75}{75}{+300}{-40}
\vskip -0.7truein
\setcounter{figure}{0}
\figcaption[f1.ps]{(b) Same as (a) except for the following. The data are 
from a 38950 s
observation with the \asca\ GIS. In the soft band image, the greyscale
runs linearly from $1.19 \times 10^{-4} \rm\, GIS\, counts \, s^{-1}\,
arcmin^{-2}$ to $2.92 \times 10^{-4} \rm\, GIS\, counts \, s^{-1}\,
arcmin^{-2}$.  In the hard band image, the greyscale runs linearly
from $2.57 \times 10^{-4} \rm\, GIS\, counts \, s^{-1}\, arcmin^{-2}$
to $5.01 \times 10^{-4} \rm\, SIS\, counts \, s^{-1}\, arcmin^{-2}$.
\label{Figure 1}}
\end{figure*}

Images (see Figure 1) were made in two energy bands, which were chosen
to optimize the signal-to-noise ratio of the detected sources. In each
of the four sets of data (i.e., the two separate GIS and SIS
detectors) two sources were apparent.  One, a soft unresolved object,
was detected with a statistical significance of 8.0$\sigma$ in the SIS
and 7.5$\sigma$ in the GIS.  This source is coincident with a
previously known X-ray source from the \eo, 2E~1552.8$-$5316, (Wilson
1986; Harris et al.~1990) and the \rosat\ all sky survey,
1RXS~J155644.7$-$532441 (Voges et al.~1999).  We will not discuss this
source any further, other than to note that it appears only in the
band below 2 keV and is therefore likely to be a relatively nearby
Galactic object.

In the band above 2 keV, at a position consistent with that of \gsnr,
we detect a new X-ray source (designated \newsrc) with a statistical
significance of 8.4$\sigma$ in the SIS and 8.5$\sigma$ in the GIS.
There are roughly $167\pm20$ SIS events and $222\pm26$ GIS events from
the new source, which is unresolved at the spatial resolution afforded
by the \asca\ telescopes (i.e., roughly $3^\prime$ half power
diameter).  There are no significant X-ray sources elsewhere in
the fields of view of either the SIS or GIS other than the two we
introduce above.

\begin{figure*}[t]
\plotfiddle{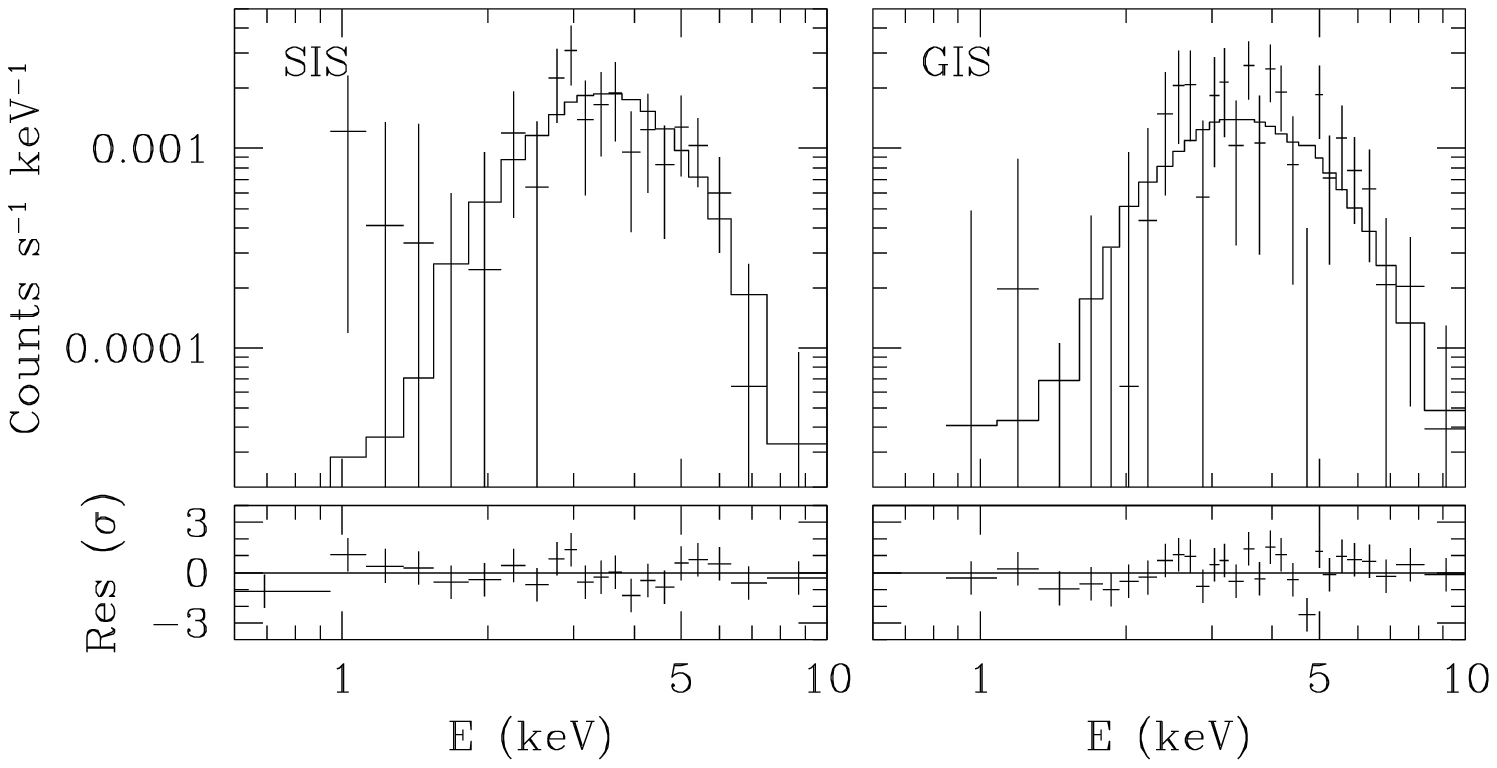}{5in}{0}{100}{100}{-300}{-50}
\vskip -1.75truein
\figcaption[f2.ps]{X-ray spectra of \gsnr\ from the \asca\ SIS and GIS
data as indicated.  The best-fit power-law spectral model is shown as
the histogram.
\label{Figure 2}}
\end{figure*}

For completeness we carried out a search for coherent pulsations in
\newsrc\ using the 2--10 keV GIS data.  As a consequence of the
re-assignment of some telemetry to timing information, as mentioned
above, the time resolution of the GIS data was 0.015625 s and 0.125 s
during the high and medium bit rate portions of the observation.
These data were separately Fourier transformed (after correcting
photon arrival times to the solar system barycenter) and the resulting
power spectra were examined for significant peaks.  The periods
searched were 0.03125 s to 1024 s for the high time resolution data
and 0.25 s to 4096 s for the medium time resolution data.  No
significant pulsed signal was detected, which is not surprising given
the small number of source photons. We set a broad limit on the pulsed
fraction of $\lsim$50\%.

Source and background spectra were extracted from both the SIS and
GIS. The source regions were circles centered on the new X-ray source
position with radii, chosen to optimize the signal-to-noise ratio of
the extracted spectra, of $3\myarcmin2$ (SIS) and $3\myarcmin9$ (GIS).
The background region for the SIS was taken from the apparently blank
portion of the field of view surrounding the source toward the NW, SW,
and SE (see Fig 1).  For the GIS background we used an annular region
centered on the position of the optical axis with inner and outer
radii chosen so that the annulus just contained the circular source
region.  This was done to ensure that the background region contained
emission that had experienced the same off-axis telescope properties
(e.g., vignetting) as the source emission. Of course, both the source
region and a region encompassing the X-ray emission from the other
soft X-ray source were excluded from the GIS background annulus.

Local background regions were used because of concerns that diffuse
emission from the well-known Galactic ridge emission (Worrall et
al.~1982) might be present in the field of view.  In order to address
this point the GIS spectra of the background regions were themselves
studied using the new (point-source removed) blank-field background
files from high Galactic latitudes. There is a clear excess of X-ray
emission in the background regions near \gsnr\ that can be fit well by
an absorbed thermal plasma model (Raymond \& Smith 1977) with $kT\sim
6-7$ keV and a surface brightness of $1.3\times 10^{-4}$ counts
s$^{-1}$ arcmin$^{-2}$ or equivalently (in terms of flux over the
2--10 keV band) $1.1\times 10^{-14}$ ergs s$^{-1}$ cm$^{-2}$
arcmin$^{-2}$. This is comparable to the brightness level expected
for the ridge near this position.  Our use of a local background is,
therefore, not only justified but essential.

Response files for spectral fitting were generated employing standard
software tools in the usual manner for \asca.  The combined SIS and
GIS spectral data were regrouped so that each spectral bin contained a
minimum of 25 events in order to allow use of $\chi^2$ as the
figure-of-merit function for spectral fitting. Given the limited
statistical quality of the data, only two simple spectral models were
explored: an absorbed power-law and an absorbed thermal plasma model
(Raymond \& Smith 1977).  The spectral data and power-law model are
shown in Figure 2 and numerical values for the best-fit parameters are
given in Table 1.  A slightly better fit was obtained for the
power-law model, although both models were statistically acceptable.
The observed 2--10 keV band flux of the source is $(6.0\pm0.8)\times
10^{-13}$ erg s$^{-1}$ cm$^{-2}$.

\begin{deluxetable}{lc}
\tablecaption{Spectral Model Fits for \gsnr}
\tablewidth{4.75truein}
\tablehead{
\colhead{Parameter} & \colhead{Value and Error (1 $\sigma$)}
}
\startdata
\multicolumn{2}{c}{Power-law Model} \nl
$N_{\rm H}$ (H atoms cm$^{-2}$) & $9.1^{+3.0}_{-2.4} \times 10^{22}$ \nl
$\alpha_{\rm P}$                & $2.9^{+0.9}_{-0.8}$ \nl
$F_{\rm E}(1\, \rm keV)$ (photon s$^{-1}$ cm$^{-2}$ keV$^{-1}$)
  & $1.8^{+6.0}_{-1.3} \times 10^{-3}$ \nl

$\chi^2$/d.o.f           &  37.6/48 \nl

\multicolumn{2}{c}{Thermal Plasma Model} \nl
$N_{\rm H}$ (H atoms cm$^{-2}$) & $7.4^{+2.2}_{-1.7} \times 10^{22}$ \nl
$kT$ (keV)    & $4.0^{+4.5}_{-1.7}$ \nl
Fractional abundance ($\odot$)  & $<$0.68 (90\% C.L.) \nl
Emission measure (cm$^{-5}$) & $1.4^{+1.6}_{-0.6} \times 10^{11}$ \nl
$\chi^2$/d.o.f           &  38.2/47 \nl
\enddata
\end{deluxetable}

\section{DISCUSSION}

The large absorption we derive from both sets of spectral fits is
fully consistent with the total Galactic column of hydrogen along the
line-of-sight to \gsnr, $N_{\rm H} \sim 10^{23}$ atoms cm$^{-2}$.  The
column can be decomposed into three parts: (1) a neutral hydrogen
column of $N_{\rm H {\sc i}} \sim 3 \times 10^{22}$ atoms cm$^{-2}$
(Wilson 1986), (2) an ionized hydrogen column of $N_{\rm H {\sc ii}}
\sim 3 \times 10^{22}$ atoms cm$^{-2}$ estimated from the
free-electron model of Taylor and Cordes (1993), and (3) a molecular
hydrogen column of $N_{\rm H_2} \sim 2 \times 10^{22}$ molecules
cm$^{-2}$ based on the CO data from Bronfman et al.~1989 and Bitran et
al.~1997 (as reported by Slane et al.~1999).  The agreement between
these measures of the total hydrogen column supports the large distance
of 17 kpc to this remnant derived from the \ion{H}{1} absorption
studies mentioned above.

\subsection{Thermal Interpretation}

The spectral fits themselves do not preclude a thermal interpretation
for the X-ray emission from \gsnr.  However there are two main
difficulties with the interpretation that cause us to disfavor it.
First and foremost is the combination of a moderately high temperature
and a lack of significant Fe-line emission.  Mean plasma temperatures
of greater than 2 keV tend to be the exception rather than the rule
for SNRs and in all cases to our knowledge are accompanied by Fe
K$\alpha$ line emission at $\sim$6.7 keV.  The second problem arises
from the inferred evolutionary state of the remnant.  It is possible
to determine the age and explosion energy from the remnant's measured
size, temperature, and emission measure in the context of similarity
solutions for SNRs in the adiabatic phase of evolution (see, for
example, Hughes, Hayashi, \& Koyama 1998).  The results are sensitive
to assumptions about the state of equilibration between electrons and
ions.  If one assumes that the temperatures of the two species are
fully equilibrated everywhere throughout the remnant, then the age and
explosion energy of \gsnr\ are estimated to be 2000--4000 yr and
$(0.9-1.8)\times 10^{51}$ erg, using values and uncertainty ranges
from Table 1 and a total size of $2\myarcmin5$ in radius (GDG).
Although these results are nicely consistent with expected values,
various lines of evidence now suggest that full equilibration of
electron and ion temperatures at supernova shock fronts is unlikely to
be the correct scenario (Laming et al.~1996, Hughes et al.~1998). If
one relaxes the assumption of full equilibration in favor of a
scenario where Coloumb collisions regulate the exchange of energy
between the species, then there are no plausible solutions for the age
and explosion energy of \gsnr\ (i.e., the inferred explosion energies
are at least two orders of magnitude larger than the canonical value
of $10^{51}$ erg).  In summary, we consider it extremely unlikely that
\gsnr\ is a young, thermally-emitting SNR.

\subsection{Nonthermal Interpretation}

Based on the radio properties of this SNR, specifically the
filled-center morphology and high degree of linear polarization of the
nebula ($\sim$20\%; GDG), a nonthermal interpretation
of the X-ray emission is most probable.  And indeed the power-law
index we find is consistent with that of other plerionic SNRs, of
which the brightest and best known is the Crab Nebula with
$\alpha_{\rm P} = 2.1$ (Toor \& Seward 1974).  The nebular luminosity
of such sources is powered by the spin-down energy loss of a
rapidly-rotating young neutron star and the emission mechanism is
synchrotron radiation from relativistic electrons in the presence of
the magnetic field of the nebula.  In some cases, like the Crab,
pulsed radiation from the central compact star is observed, although
in other cases it is not, perhaps due to unfavorable viewing geometry
of the presumably beamed radiation from the pulsar.  However, emission
from the synchrotron nebula is isotropic and therefore, when detected,
a synchrotron nebula is believed to provide clear evidence for the
presence of a young pulsar.

\begin{figure*}[b]
\plotfiddle{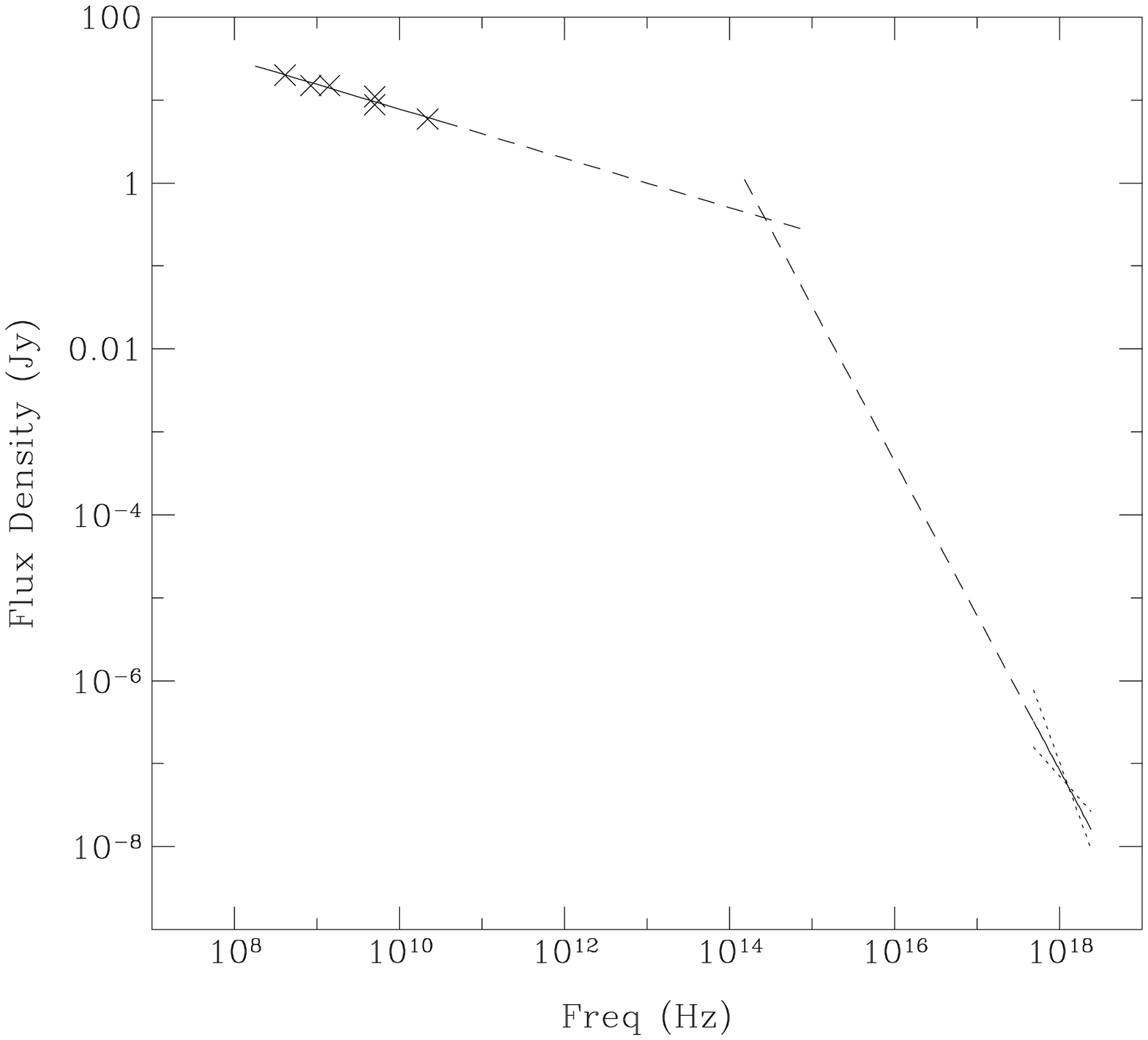}{3.5in}{0}{50}{50}{-150}{-100}
\figcaption[f3.ps]{Broadband spectrum of \gsnr\ from radio to X-ray
frequencies. Radio data points come from Caswell et al.~(1975) (1.42
GHz); Clark \& Caswell (1976) (0.408 GHz and 5 GHz).  Tateyama,
Sabalisck, \& Abraham (1988) (5 GHz and 22 GHz); and Whiteoak \& Green
(1996) (0.843 GHz). The dashed curves show the extrapolation of the
best-fit radio and X-ray power-law models to the point where they
intersect at the so-called break frequency, which is
$\sim$$3\times10^{14}$ Hz (albeit with large errors).
\label{Figure 3}}
\end{figure*}

The unabsorbed soft X-ray luminosity (0.2--4 keV band) of \gsnr\ is
$\log(L_X/\rm erg \, s^{-1}) = 35.6^{+1.0}_{-0.7}$. This value is
larger than all known pulsar-powered synchrotron nebula with the
exception of the Crab 
%
%
(Seward \& Wang 1988),
and two SNRs in the Large Magellanic Cloud: E0540$-$69.3
%
%
(Seward \& Wang 1988)
and N157B
%
%
(Wang \& Gotthelf 1998).  
It is comparable to the luminosities of the
Crab-like cores of G29.7$-$0.3 (Blanton \& Helfand 1996), G21.5$-$0.9
(Slane et al.~2000), and MSH 15$-$5{\sl 2} (Seward \& Wang 1988).
From the empirical relationship between the soft X-ray luminosity of
the synchrotron nebula, $L_X$ and the pulsar's rate of loss of
rotational energy, $\dot{E}$ (Seward \& Wang 1988), we determine that
the (still undiscovered) pulsar that must be powering the synchrotron
emission is losing rotational energy at the rate of $\dot{E} \sim
10^{37} - 2\times10^{38} \rm erg\, s^{-1}$.  GDG use the radio
luminosity to estimate a rotational energy loss rate of several
$10^{38}\, \rm erg\, s^{-1}$.  Given the large scatter in the
correlations between $\dot{E}$ and luminosity in each band, the
differences between these estimates are unlikely to be significant.
Indeed, focusing on the more interesting broad agreement between them,
we conclude that \gsnr\ harbors a rapidly spinning pulsar that is
losing rotational energy at a rate of $\sim$$10^{38}\, \rm erg\,
s^{-1}$.

GDG argue that the size and radio luminosity of \gsnr\ most closely
resemble those of N157B, the SNR that contains the most rapidly
spinning known pulsar with a period of 0.016 s (PSR 0537$-$6910;
Marshall et al.~1998).  This is also the case for the X-ray luminosity
and for the ratio of X-ray to radio luminosity. The X-ray luminosity
of the power-law component in N157B is $\log(L_X/\rm erg \, s^{-1}) =
36.7$, at the upper end of our allowed range for \gsnr.  Although the
ratio of X-ray to radio luminosity is low for \gsnr, $L_X/L_R =
0.3-25$, nevertheless it is an excellent match to that of N157B
($L_X/L_R \sim 10$).  In contrast $L_X/L_R$ are 130 and 800 for the
Crab and E0540$-$69.3, respectively. According to Wang \& Gotthelf
(1998) the X-ray synchrotron nebula in N157B is some $5\times7$ pc in
size and is surrounded by a fainter diffuse component of thermal X-ray
emission from the SN blast wave. Because of the broad
point-spread-function of \asca, we are unable to place interesting
limits on the size of the X-ray synchrotron nebula in \gsnr. This must
await more sensitive and higher spatial resolution X-ray
observations. Unfortunately, the large absorption toward \gsnr\ may
forever preclude the detection of soft thermal X-ray emission (with
$kT\sim 1$ keV) from any outer shell or blast wave in \gsnr.  We note
that there is no evidence for a shell from the high spatial resolution
radio image (GDG).

The broadband spectrum of \gsnr\ from radio to X-ray frequencies is
shown in Figure 3. The point at which the extrapolated best-fit radio
and X-ray power-laws intersect, the so-called break frequency, is
$\nu_{\rm B} \sim 3\times10^{14}$ Hz.  The large uncertainty on the
measured X-ray spectral index results in a huge range for $\nu_{\rm
B}$ ($10^{11}$ Hz -- $6\times10^{15}$ Hz) that provides little more
than consistency with values from other SNRs, e.g., the Crab with
$\nu_{\rm B} \sim 10^{13}$ Hz.  Compared to the Crab, \gsnr\ is nearly
as luminous in the radio band (GDG), while it is some 70 times less
luminous in X-rays.

Synchrotron radiation theory (e.g., Ginzburg \& Syrovatskii 1965)
provides a means for estimating from the observed spectrum the amount
of energy in relativistic particles and magnetic field in the nebula.
One needs an additional constraint which we take to be the condition
that the total energy is at a minimum. In this case the magnetic field
energy is 3/4 of the energy in particles.  We further assume that the
energy in all relativistic particles is twice the energy in the
electron component alone.  We integrate the broadband spectrum
(Fig.~3) from low frequency to $\nu_{\rm B}$ using the radio spectral
index and from $\nu_{\rm B}$ to $2.4\times 10^{18}$ Hz (10 keV) using
the X-ray spectral index.  For our derived quantities we show
explicitly the dependencies on distance, $D$, angular radius of the
nebula, $\theta$, the ratio of the energy in magnetic field to that in
particles, $\kappa_m$, and the ratio of energy in all particles to
electrons, $\kappa_r$. We use a value, $\theta = 1^\prime$, for the
size of the nebula that corresponds to the bright central portion
(GDG).  The value we derive for the nebular magnetic field is $ B =
365\, (\kappa_m/0.75)^{2/7} (\kappa_r/2)^{2/7} (D/{17\, \rm
kpc})^{-2/7} (\theta/1^\prime)^{-6/7}$ $\mu\rm G $ and the energy in
relativistic electrons is $ W_e = 5\times 10^{49}\, (D/{17\, \rm
kpc})^2 (B/365\, \mu \rm G)^{-3/2} \, \rm erg$, with roughly equal
amounts of energy in the populations of electrons above and below the
break frequency.  Our estimates of the above quantities are reduced to
$B \sim 220\,\mu\rm G$ and $W_e \sim 2\times 10^{49} \,\rm erg$ if we
use the lower limit on the break frequency ($10^{11}$ Hz) that comes
from the 1-$\sigma$ lower limit on the X-ray spectral index.  When the
total energy in the nebula is computed, viz., $W = W_{cr} + W_{B} =
\kappa_r (1+\kappa_m) W_e$, even the lower limit value is still rather
high, $W = 7 \times 10^{49} \,\rm erg$.

Although the energy we see currently in the synchrotron nebula must
have come from the initial rotational energy of the pulsar, $E_0 =
2\pi^2I P_0^{-2}$, it is also clear that $W$ can represent only a
fraction of $E_0$.  Some of the initial rotational energy of the
pulsar has gone into radiation and the expansion of the nebula and
some still remains in the form of the current spin of the pulsar.
Calculating the precise partition of $E_0$ into the various forms that
it may take is a challenging task and, given the limited quality of
the current data, is beyond the scope of this work.  Nevertheless by
equating $W$ to $E_0$ we can derive a strong upper limit to the
initial spin period of the pulsar.  Using $I=10^{45}\,\rm g\, cm^2$
for the moment of inertia, the limit we obtain is $P_0 < 0.017\,\rm
s$.  This limit is fully consistent with the current spin period of
PSR 0537$-$6910 and adds further weight to the preceding arguments, as
well as those in GDG, that the closest known analogue to \gsnr\ is the
Crab-like SNR N157B.

\section{CONCLUSIONS}

Using \asca\ we have discovered hard X-ray emission (2-10 keV band)
from the Galactic radio SNR \gsnr.  The new X-ray spectral data can be
equally well fit by highly absorbed ($N_{\rm H} \sim 10^{23} \, \rm
atoms\, cm^{-2}$) thermal plasma or power-law models. When the
inferred astrophysical properties of the remnant under each scenario
are examined, along with additional constraints from recent radio
observations, we conclude that \gsnr\ is a Crab-like SNR dominated by
synchrotron emission in the radio and X-ray bands.  It is powered by
the spin-down energy of a central, yet undetected, pulsar that is
currently losing energy at a rate of $\sim$$10^{38}$ erg s$^{-1}$.  In
order to account for the total energy in particles and magnetic field
observed in the synchrotron nebula at the present time, we find that
the initial spin period of the pulsar must have been no more than
0.017 s.  In many respects \gsnr\ bears a particularly close
resemblance to the Crab-like SNR N157B in the Large Magellanic Cloud,
the remnant containing the most rapidly rotating pulsar known, PSR
0537$-$6910.  The discovery of its hidden pulsar is clearly an
important next step in furthering our understanding of \gsnr, yet the
faintness of the remnant in the X-ray band ($6\times 10^{-13}$ erg
s$^{-1}$ cm$^{-2}$ over 2--10 keV) will make this search a challenging
task.

\acknowledgements

This work made use of the SIMBAD Astronomical Database maintained by
the Centre de Donn\'ees Astronomiques de Strasbourg.  Useful
conversations with Cara Rakowski and Brian Gaensler are greatly
acknowledged.  We appreciate Monique Arnaud's support and hospitality
during the course of this project.  Partial support was provided by
NASA grant NAG5-6420 (JPH) and contract NAS8-39073 (POS, PPP).

\end{document}